# PorousGen: An Efficient Algorithm for Generating Porous Structures with Accurate Porosity and Uniform Density Distribution

Shota Arai[1] & Takashi Yoshidome[1]

[1]*Department of Applied Physics, Graduate School of Engineering, Tohoku University, Sendai 980-8579, Japan*

Corresponding Author: Shota Arai, shota.arai.c2@tohoku.ac.jp

ABSTRACT

This work presents a novel algorithm for generating porous structures as an alternative to the *PoreSpy* program suite. Unlike *PoreSpy*, which often produces structures whose porosity deviates from the target value, our proposed algorithm generates structures whose porosity closely matches the specified input, within a defined error margin. Furthermore, parallel computation enables efficient generation of large-scale structures, while memory usage is reduced compared to *PoreSpy*. To evaluate performance, structures were generated using both *PoreSpy* and the proposed method with parameters corresponding to X-ray ptychography experiments. The porosity mismatch in *PoreSpy* led to a relative error exceeding 20% in the computed gas diffusion coefficients, whereas our method reproduced the experimental values within 5%. These results demonstrate that the proposed method provides an efficient, high-precision approach for generating porous structures and supports reliable prediction of material properties. The program called "PorousGen" is publicly available under the MIT License from https://github.com/YoshidomeGroup-Hydration/PorousGen.

## 1  Introduction

Computational methods for generating porous structures have widely applied in various research fields. For example, in the field of digital rock physics, the permeability of porous structures has been computed using thin-section images [1,2]. In addition, artificially generated datasets of porous structures have been used to efficiently predict permeability [3,4] and gas diffusion coefficients [5-12] . Moreover, computationally generated porous structures have also been used for the construction of machine-learning models for the structure-property relationship, compensating for the limited availability of experimental data. For example, we recently established a structure–property relationship between porous structures and their gas diffusion coefficients using manifold learning [13], which successfully predicted the diffusion coefficient of a porous structure obtained from X-ray ptychography experiments. These studies demonstrated usefulness of using computational methods for generating porous structures.

A program for generating porous structures is *PoreSpy* program suite [14]. It generates porous structures with a given particle-size distribution. However, an issue arises when generating porous structure with polydisperse spheres. As described in the web page of polydisperse_spheres function [15,16], because the porosity is only matched approximately, and thus the check of the porosity value of the generated porous structure is necessary. This approximate matching limits accurate computation of physical quantities.

In this study, we propose a new porous structure generation algorithm that differs fundamentally from PoreSpy. This algorithm not only resolves the limitations identified in the previous section but also offers two major advantages. First, it ensures that the generated structures achieve the target porosity with high accuracy by iteratively adjusting until the input and generated porosities match within a specified error range. Moreover, it maintains spatially uniform porosity, overcoming the tendency of PoreSpy to exhibit visually higher porosity near the outer boundaries compared with the central region. Second, our algorithm achieves higher computational efficiency, requiring less computation time through parallelization and lower memory consumption. Overall, it demonstrates both high accuracy and efficiency in porous structure generation.

## 2 Methods

### 2.1. Proposal of an algorithm for generating a porous structure

We developed the following algorithm for the generation of a porous structure with the target porosity of $\eta_{\text{Box}}$ (Fig. 1):

1. Preparation of a box:
   A box with the side length of $(L_{\text{Box}} + l)$ is prepared (Fig. 1(a)). To suppress the artificially high porosity near the boundaries, only the central portion with side length $L_{Box}$ in step 5 is later extracted (step 5). Periodic boundary conditions are not applied here because real porous structures do not satisfy such periodicity. In sharp contrast, *PoreSpy* neither applies periodic boundary conditions nor implements a central portion extraction step. Consequently, simulations utilizing *PoreSpy* often exhibit an artifactual increase in porosity near the box edges, as discussed in the "Results and Discussion" section.

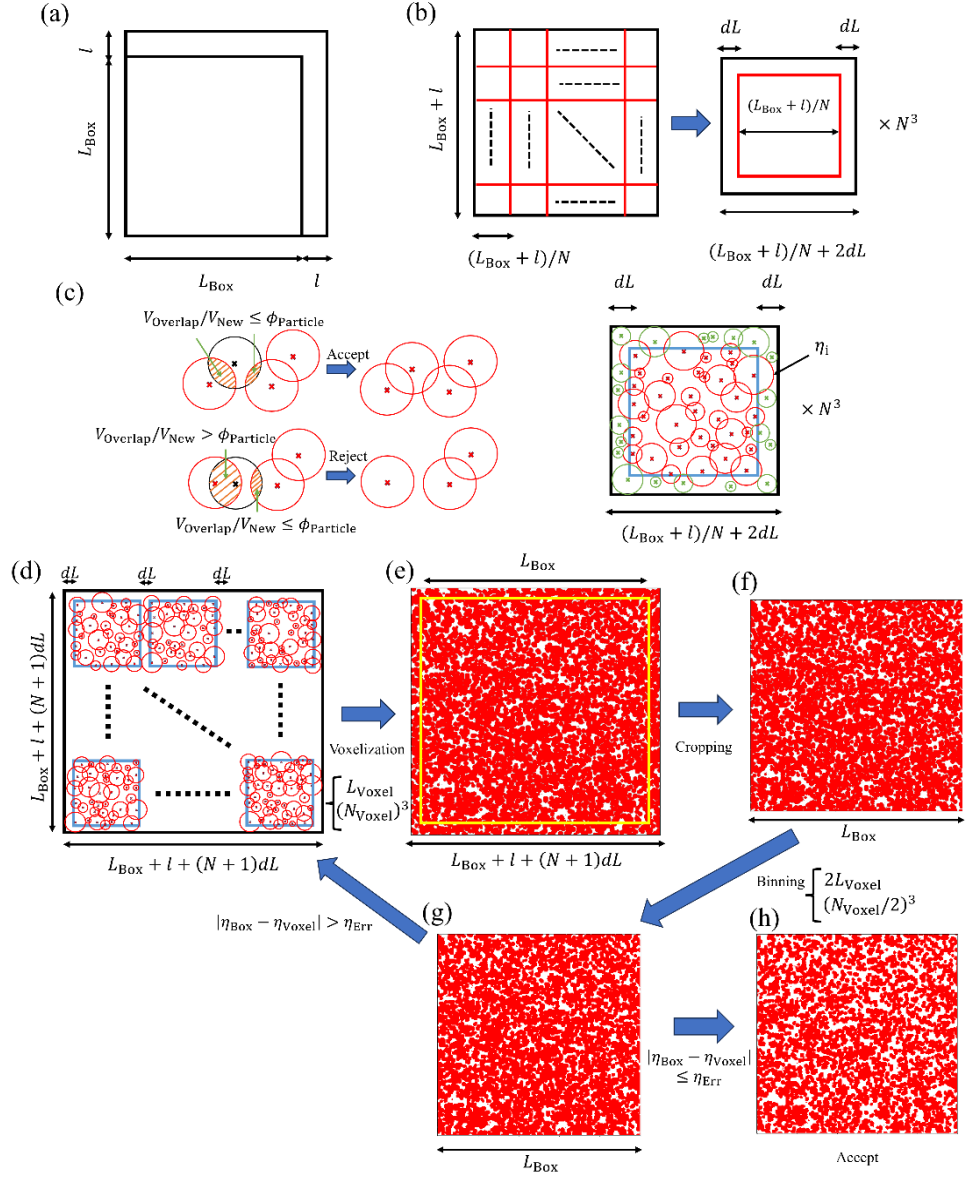

**Figure 1** Schematic of the generation process. (a) Initial system setup with side length $(L_{Box} + l)$. (b) System divided into $N^3$ subdomains, each of which extended by $2dL$. Red square indicates the original system. (c) Particles are placed in each subdomain until the porosity $\eta_{Box}$ is reached. Red particles have centers inside the original subdomain, while green particles have centers in the extended region (removed afterward). (d) Reconstruction by overlapping subdomains so that the region of $dL$ matches. (e) Voxelization with the voxel size $(L_{Voxel})^3$ and total number of voxels $(N_{Voxel})^3$. The yellow square indicates the cropping region. (f) Cropped system of size $L_{Box}$. (g) the map is binned, generating the map with the voxel size $(2L_{Voxel})^3$ and the total number of voxels $(N_{Voxel}/2)^3$. (h) If $|\eta_{Box} - \eta_{Voxel}| \leq \eta_{Err}$, the system is accepted.

2. Subdivision and expansion of the box for parallel computation:
   The box was divided into $N^3$ subdomains for parallel computation (Fig. 1(b)). Each subdomain was further expanded by $2dL$ along each side to suppress artifacts at the boundaries when merging the subdomains in step 4.

3. Particle placement:
   Particles were placed in each subdomain according to the specified placement rules. For each small box, particles were placed until the porosity reached $\eta_{\text{Box}}$ (Fig. 1(c)). The coordinates of particles were chosen randomly, and their diameters were determined according to an arbitrarily specified distribution with mean and standard deviation value $D_{\text{Mean}}$ and $D_{\text{SD}}$. A new particle was placed only if the overlap ratio $\phi_{\text{Particle}}$ between particles satisfied the following condition:

   $$V_{\text{Overlap}}/V_{\text{New}} \leq \phi_{\text{Particle}}, \tag{1}$$

   where, $V_{\text{New}}$ is the volume of the newly placed particle, and $V_{\text{Overlap}}$ is the volume overlapping with an already placed particle, calculated as

   $$V_{\text{Overlap}} = \frac{\pi}{12d}(r_{\text{New}} + r_{\text{Old}} - d)^2 \{d^2 + 2d(r_{\text{New}} + r_{\text{Old}}) - 3(r_{\text{New}} - r_{\text{Old}})^2\}. \tag{2}$$

   Here, $r_{\text{New}}$ and $r_{\text{Old}}$ are the radii of the new and existing particles, respectively, and $d$ is the distance between their centers. If $d \geq r_{\text{New}} + r_{\text{Old}}$, then $V_{\text{Overlap}} = 0$. Conversely, if $d \leq |r_{\text{New}} - r_{\text{Old}}|$, $V_{\text{Overlap}}$ was taken as the volume of the smaller particle. Furthermore, particles whose centers were located within a distance $dL$ from the box boundary were excluded to reduce boundary effects after merging (Fig. 1(c)). In addition, overlaps between particles and the edges of the sub-box were prohibited.

4. Subdomain merging:
   After all subdomains were filled with particles, they were merged using the overlapping $dL$ regions. Following the merge, the side length of the system became $L_{\text{Box}} + l + (N + 1)dL$ (Fig. 1(d)).

5. Voxelization, trimming, and binning:
   The merged system was converted to a map by assigning a value of 1 to voxels containing a particle and 0 to others (Fig. 1(e)). The side length of each voxel was $L_{\text{Voxel}}$ and the total number of voxels was $(N_{\text{Voxel}})^3$. Next, the central region was cropped to have a side length of $L_{Box}$ (Fig. 1(f)). Then, binning was applied to reduce the influence on the power spectrum (Fig. 1(g), [13], and Fig. S1 in the supporting information) by using the block_reduce function from the *scikit-image* library [17] to average the intensity values of adjacent 2×2×2 voxels. The resultant map consists of $(N_{Voxel}/2)^3$ voxels with a voxel size of $2L_{Voxel}$. After the binning, the median intensity value was set as a threshold, and all intensity values below this threshold were set to zero.

6. Porosity adjustment

    The generated system was accepted if $|\eta_{Box} - \eta_{Voxel}| \leq \eta_{Err}$, where $\eta_{Voxel}$ is the porosity of the generated system, and $\eta_{Err}$ is the tolerance (Fig. 1(h)). If this criterion was not satisfied, the system was reverted to the particle configuration obtained at step 4, and the following particle insertion or removal was conducted: If $\eta_{Voxel} < \eta_{Box}$, a particle was removed if a uniform random number $r_i \in [0,1]$ exceeded, $p_i = (1 - \eta_{Box})/(1 - \eta_{Voxel})$; Conversely, if $\eta_{Voxel} > \eta_{Box}$, a particle with a radius of $r$ drawn from the prescribed distribution was randomly inserted when satisfying the condition of eq. (1). After the insertion or removal, the step 5 was conducted to obtain a new map, and the steps 4 and 5 were repeated until the condition of $|\eta_{Box} - \eta_{Voxel}| \leq \eta_{Err}$ was reached.

## 2.2. Generation of porous structures using the present algorithm

In this study, the particle diameters used for structure generation follow a log-normal distribution. In all calculations, the tolerance $\eta_{Err}$ was fixed at 1%. The other parameters are listed in Table 1. The "–" indicates that the parameter is treated as a variable.

It is noted that the $\eta_{Box}$ value shown in Table 1(d) arises from the $\eta_{Voxel}$ value of *PoreSpy* obtained with the input of the $\eta_{Box}$ value of 50 % (Other parameters required for the *PoreSpy* were the same as those shown in Table 1(d)). Because of the issue described in the Introduction section, the input porosity value ($\eta_{Box} = 50$ %) was slightly different from the porosity value of the generated porous structure ($\eta_{Voxel} = 46.2$ %). To compare the generated maps between the *PoreSpy* and our algorithm, $\eta_{Box}$ value was set to 46 % in our algorithm (It is noted that in our algorithm the difference between $\eta_{Box}$ and $\eta_{Voxel}$ in our algorithm is less than 1 % because of the $\eta_{Err}$ value of 1%).

Table. 1 Parameters used for the present algorithm

| $\eta_{Box}$ | $L_{Box}$ | $L_{Voxel}$ | $N_{Voxel}$ | $D_{Mean}$ | $D_{SD}$ | $N$ | $l$ | $2dL$ [nm] | $\phi_{Particle}$ |
|---|---|---|---|---|---|---|---|---|---|

|     | [%] | [μm] | [nm] |      | [nm] | [nm] |   | [μm] |                          | [%] |
|-----|-----|------|------|------|------|------|---|------|--------------------------|-----|
| (a) | 50  | 2    | 5    | 400  | 60   | 1    | 5 | —    | $D_{\text{Mean}} + 2D_{\text{SD}}$ | 50  |
| (b) | 50  | 2    | 5    | 400  | 60   | 1    | 5 | 0.25 | —                        | 50  |
| (c) | 50  | 2    | 5    | 400  | 60   | 1    | 5 | 0.25 | $D_{\text{Mean}} + 2D_{\text{SD}}$ | —   |
| (d) | 46  | 10   | 20   | 500  | 140  | 1    | 5 | 1    | $D_{\text{Mean}} + 2D_{\text{SD}}$ | 50  |
| (e) | 34  | 5    | 5    | 1000 | 40   | 5    | 6 | 0.25 | $D_{\text{Mean}} + 2D_{\text{SD}}$ | 50  |

### 2.3. Generation of porous structures using the *PoreSpy* program suite

To compare the results of our algorithm with those of *PoreSpy*, we generated three-dimensional electron density maps using the parameters shown in Tables 1(d) and 1(e), except for the setting $\eta_{\text{Box}} = 50$ % because of the reason described in the previous subsection. The same distribution as that employed in our algorithm (log-normal distribution) was used.

### 2.4. Computation of physical properties

For the maps generated by our algorithm and by *PoreSpy*, the diffusion coefficients were calculated using the ps.simulations.tortuosity_fd function of the *PoreSpy* library [18]. This function computes the tortuosity of a map, and the gas diffusion coefficient was obtained as its inverse. The linear systems arising from the finite-difference-based simulations were solved using the cg solver from the *SciPy* library [19]. Because tortuosity in *PoreSpy* is dimensionless, the gas diffusion coefficient derived from it is also dimensionless.

Similarly, the actual porosity of the numerically measured map was calculated using the same function, excluding pores that do not contribute to diffusion. Hereafter, this porosity is referred to as the effective porosity. In the present calculations, regardless of whether *PoreSpy* or our algorithm was used, the difference between the effective porosity and $\eta_{\text{Box}}$ was always less than 1 %.

### 2.5. Computation of power spectrum

First, the three-dimensional Fourier transform of the three-dimensional electron density map was performed using the fftn function from the *SciPy* library, which implements the Fast Fourier Transform (FFT). Next, the power spectrum $S(\mathbf{k})$ was obtained as $S(\mathbf{k}) = F(\mathbf{k})F^*(\mathbf{k})$, where $F(\mathbf{k})$ is the Fourier transform of the three-dimensional electron density map.

### 2.6. Experimental data

To compare with an X-ray ptychography experiment data [13], maps were generated using our algorithm and *PoreSpy* with the parameters shown in Table 1(e). The diffusion coefficient and effective porosity were then calculated for these maps. According to [13], the dimensionless diffusion coefficient and effective porosity of the experimental data are $33.8$ and $0.087$, respectively.

## 3 Results and Discussion

### 3.1. Parameter dependence of our algorithm

#### 3.1.1. Dependence on the parameter $l$

We first investigated the effect of the parameter $l$ in our algorithm on the generated maps (Fig. 1(a)). The parameters listed in Table 1(a) were used. The ranges of the maps were $0\ \mu m \leq X \leq 2.0\ \mu m$, $0\ \mu m \leq Y \leq 2.0\ \mu m$, and $0\ \mu m \leq Z \leq 2.0\ \mu m$.

The generated maps and their corresponding power spectra are shown in Fig. 2. In the maps generated with $l = 0\ \mu m$, the porosity near the outer boundary ($Z = 0\ \mu m$, Fig. 2(a)) appeared visually higher than that at the central region ($Z = 5\ \mu m$, Fig. 2(b)). This non-uniformity near the boundaries caused cross-shaped peaks in the power spectrum (Fig. 2(c)). The origin of this effect lies in the constraint that particles are not allowed to extend beyond the system boundaries, which reduces particle density near the edges (Figs. 1(a) and 1(e)).

In contrast, when our algorithm was applied with $l = 0.25\ \mu m$, the generated maps exhibited uniform porosity regardless of location (Figs. 2(d) and 2(e)). This uniformity is a result of the preprocessing step implemented to eliminate boundary effects (Figs. 1(a) and 1(e)). Consequently, the influence of the boundaries on the maps and their power spectra was minimized (Fig. 2(f)).

#### 3.1.2. Dependence on the parameter $dL$

We next investigated the effect of the parameter $dL$ in our algorithm on the generated maps (Fig. 1(c)). The parameters listed in Table 1(b) were used. The ranges of the maps were $0\ \mu m \leq X \leq 2.0\ \mu m$, $0\ \mu m \leq Y \leq 2.0\ \mu m$, and $0\ \mu m \leq Z \leq 2.0\ \mu m$.

The resultant three-dimensional electron density maps and the corresponding power spectra are shown in Fig. 3. When maps were generated with $dL = 0\ \mu m$, the porosity depended on the position: While grid-like voids were observed in the $X - Y$ cross-section at $Z = 0.15\ \mu m$ (Fig. 3(a)), the porosity tended to be higher in the $X - Y$ cross-section at $Z = 0.30\ \mu m$ (Fig. 3(b)). This effect also appears as cross-shaped peaks in the power spectrum (Fig. 3(c)). The reason of such low density at the boundaries arises from the prohibition of overlaps between particles and the boundaries.

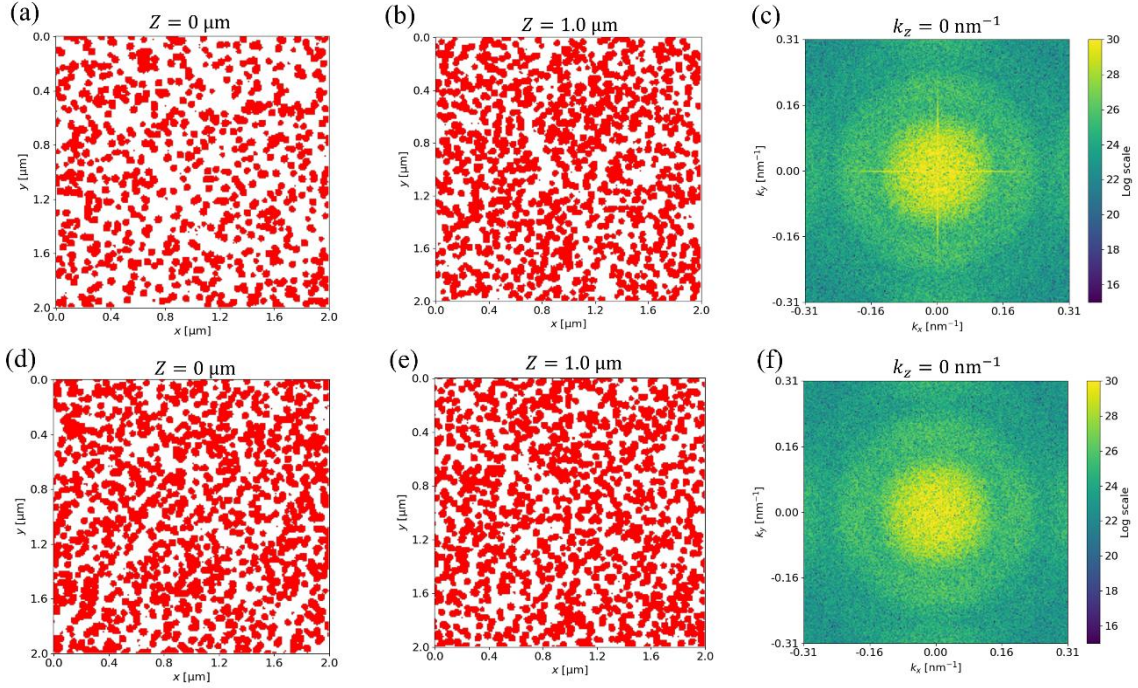

**Figure 2** Comparison of maps generated by our algorithm with different lengths $l$. The ranges of the maps were $0\ \mu m \leq X \leq 2.0\ \mu m$, $0\ \mu m \leq Y \leq 2.0\ \mu m$, and $0\ \mu m \leq Z \leq 2.0\ \mu m$. (a, b) the three-dimensional map generated with $l = 0\ \mu m$ at $Z = 0\ \mu m$ and $1.0\ \mu m$. (c) Power spectrum of the map generated with $l = 0\ \mu m$ at $k_z = 0\ nm^{-1}$. (d, e) the three-dimensional map generated with $l = 0.25\ \mu m$ at $Z = 0\ \mu m$ and $1.0\ \mu m$. (f) Power spectrum of the map generated with $l = 0.25\ \mu m$ at $k_z = 0\ nm^{-1}$. The other parameters are listed in Table 1(a).

In contrast, when $2dL = D_{\text{Mean}}$, the generated maps exhibited uniform porosity regardless of location (Figs. 3(d) and 3(e)) and the influence of the boundaries on the power spectrum was minimized (Fig. 3(f)). Because visually identical results were obtained when $2dL = D_{\text{Mean}} + 2D_{\text{SD}}$ (Figs. 2(d)–2(f)) in our algorithm, setting $2dL$ to a value equal to or greater than $D_{\text{Mean}}$ effectively suppresses artifacts near the subdivided regions after merging (Fig. 1(d)).

### 3.2.3. Dependence on the parameter $\phi_{\text{Particle}}$

We also investigated the effect of the parameter $\phi_{\text{Particle}}$ in our algorithm on the generated maps (Fig. 1(c)). The parameters listed in Table 1(c) were employed. The ranges of the maps were $0\ \mu m \leq X \leq 2.0\ \mu m$, $0\ \mu m \leq Y \leq 2.0\ \mu m$, and $0\ \mu m \leq Z \leq 2.0\ \mu m$.

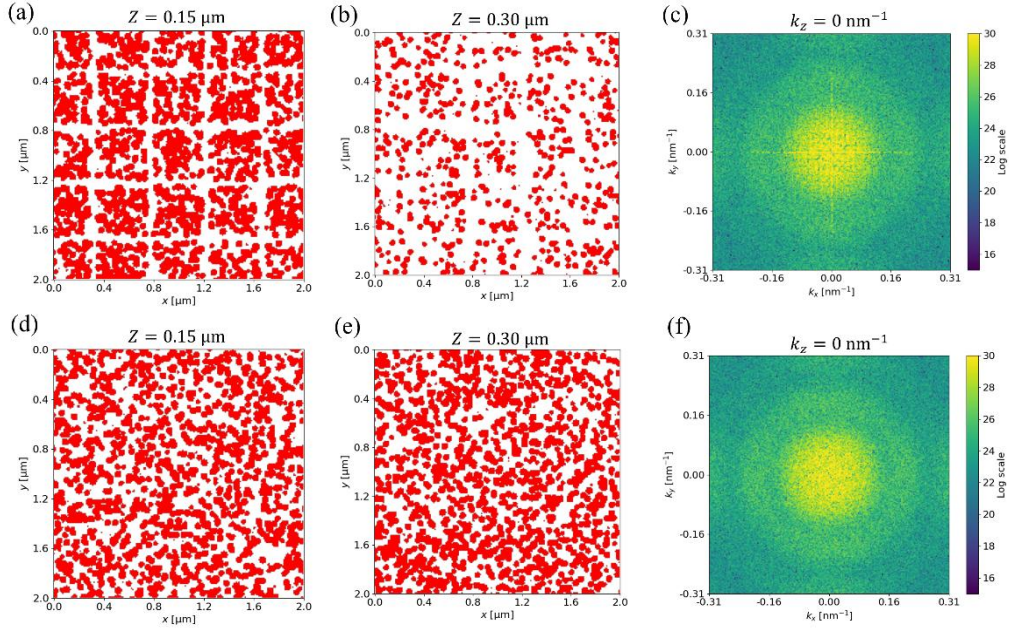

**Figure 3** Comparison of maps generated by our algorithm with different lengths $2dL$. The ranges of the maps were $0\,\mu m \leq X \leq 2.0\,\mu m$, $0\,\mu m \leq Y \leq 2.0\,\mu m$, and $0\,\mu m \leq Z \leq 2.0\,\mu m$. (a, b) $X-Y$ cross-sections at $Z = 0.15\,\mu m$ and $Z = 0.30\,\mu m$ of the three-dimensional map generated with $2dL = 0\,\mu m$. (c) Power spectrum of the map generated with $2dL = 0\,\mu m$ at $k_z = 0\,nm^{-1}$. (d, e) $Y-Z$ cross-sections at $Z = 0.15\,\mu m$ and $Z = 0.30\,\mu m$ of the three-dimensional map generated with $2dL = D_{Mean}$. (f) Power spectrum of the map generated with $2dL = D_{Mean}$ at $k_z = 0\,nm^{-1}$. The other parameters are listed in Table 1(b).

Varying $\phi_{Particle}$ to $25\,\%$, $50\,\%$, $75\,\%$, and $100\,\%$ did not result in significant differences in the three-dimensional porous maps or their power spectra (Fig. 4). Additionally, the computational time and memory usage remained nearly the same, comparable to the cases with $25-100\,\%$ (Data is not shown). The $0\,\%$ case, where particle overlap was completely prohibited, was not computed due to the extremely long calculation time.

### 3.2. Comparison of maps obtained using our algorithm and *PoreSpy* program suite

We generated maps using our algorithm and the *PoreSpy* program suite (Fig. 5) to compare the obtained maps. The ranges of the maps were $0\,\mu m \leq X \leq 10.0\,\mu m$, $0\,\mu m \leq Y \leq 10.0\,\mu m$, and $0\,\mu m \leq Z \leq 10.0\,\mu m$. The parameters shown in Table 1(d) were employed, except for the setting $\eta_{Box} = 50\,\%$ in the PoreSpy (see Sec. 2.3). In addition, since *PoreSpy* does not perform binning on the generated structures, the binning step was also omitted in our algorithm (Fig. 1(g)).

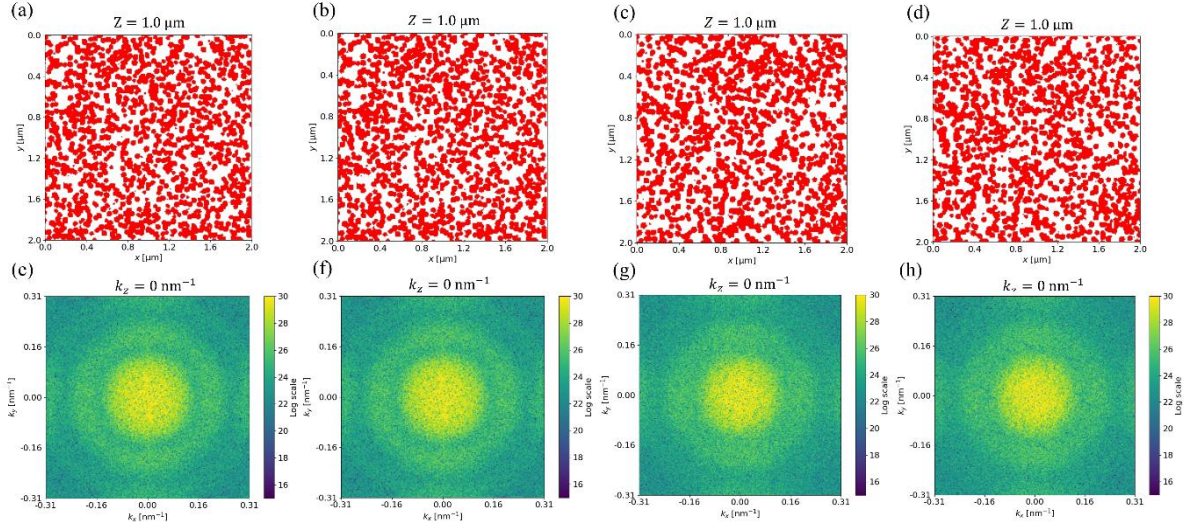

**Figure 4** Comparison of maps generated by our algorithm with different $\phi_{Particle}$ values. The ranges of the maps were $0\ \mu m \leq X \leq 2.0\ \mu m$, $0\ \mu m \leq Y \leq 2.0\ \mu m$, and $0\ \mu m \leq Z \leq 2.0\ \mu m$. (a–d) Cross-sectional views of the three-dimensional maps at $Z = 1.0\ \mu m$, corresponding $\phi_{Particle}$ to $25\ \%$, $50\ \%$, $75\ \%$, and $100\ \%$, respectively. (e–h) Power spectra of the maps shown in (a–d) at $k_z = 0\ nm^{-1}$. The other parameters are listed in Table 1(c).

It was found from Figure 5 that the maps generated by our algorithm and *PoreSpy* showed almost identical results except near the outer boundaries. In the maps generated by *PoreSpy*, the porosity near the outer boundary ($Z = 0\ \mu m$, Fig. 5(a)) appeared visually higher than that at the central region ($Z = 5\ \mu m$, Fig. 5(b)). The influence of the difference in the porosity at the boundary appeared as cross-shaped peaks in the power spectrum (Fig. 5(c)). In contrast, the maps generated by our algorithm exhibited uniform porosity regardless of location (Figs. 5(d) and 5(e)), due to the preprocessing step applied in advance to eliminate boundary effects (Figs. 1(a) and 1(e)). As a result, the boundary influence was minimized, and its effect on the power spectrum was also reduced (Fig. 5(f)).

### 3.3. Comparison with an experimental data

We compared the ability of *PoreSpy* and our proposed algorithm to reproduce a map observed in X-ray ptychography experiment [13]. The parameters are listed in Table 1(e). In the case of *PoreSpy*, binning was applied to the generated maps through the same procedure as that performed in our algorithm. Therefore, the final voxel size and the number of voxels were the same as those in the X-ray ptychography experiment and our algorithm, i.e., $10\ nm$ and $(500)^3$, respectively.

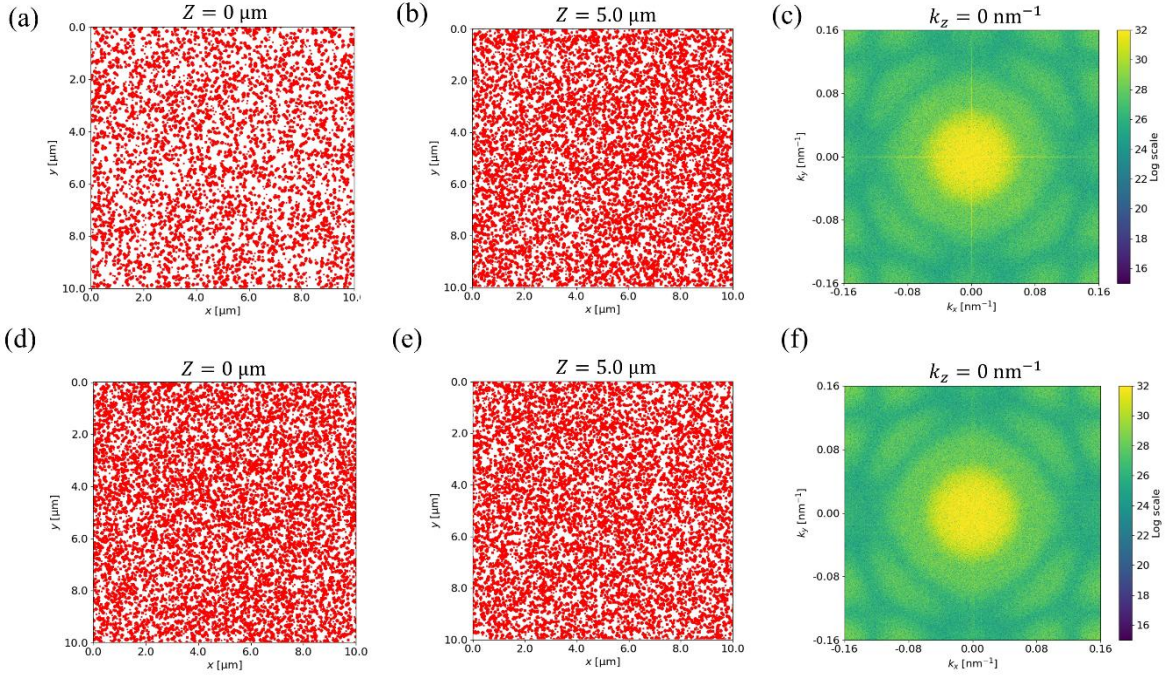

**Figure 5** Comparison of maps generated by PoreSpy and our algorithm. The ranges of the maps were $0\,\mu m \leq X \leq 10.0\,\mu m$, $0\,\mu m \leq Y \leq 10.0\,\mu m$, and $0\,\mu m \leq Z \leq 10.0\,\mu m$. (a, b) The three-dimensional map using PoreSpy at $Z = 0\,\mu m$ and $5.0\,\mu m$. (c) Power spectrum of the map by PoreSpy at $k_z = 0\,nm^{-1}$. (d, e) Maps generated by our algorithm at $Z = 0\,\mu m$ and $5.0\,\mu m$. (f) Power spectrum of the generated map by our algorithm at $k_z = 0\,nm^{-1}$. In PoreSpy, $\eta_{Box}$ was set to $50\,\%$, whereas in our algorithm it was set to $46\,\%$. The other parameters are listed in Table 1(d). Furthermore, in our algorithm, the binning step shown in Fig. 1(g) was omitted during structure generation. In our algorithm, the binning step shown in Fig. 1(g) was omitted during structure generation.

The effective porosity and the gas diffusion coefficient were summarized in Table 2. The difference in effective porosity between the X-ray ptychography experiment and *PoreSpy* was 4.3%, which resulted in a relative error of more than 20% in the gas diffusion coefficient. This indicates that even a few percent deviation in effective porosity has a significant impact on the diffusion coefficient. In contrast, the difference in effective porosity between the X-ray ptychography experiment and our algorithm was within 1%, and the relative error in the gas diffusion coefficient was only about 4.3%. Because our algorithm minimizes the deviation between $\eta_{Box}$ and the effective porosity, it was able to reproduce the experimental diffusion coefficient with high accuracy.

Table 2: Comparison of effective porosity and gas diffusion coefficients for X-ray ptychography experiment, *PoreSpy*, and our algorithm.

|  | Effective porosity [%] | Gas diffusion coeffct |
|---|---|---|
| Experiment | 33.8 | 0.087 |
| *PoreSpy* | 29.5 | 0.067 |
| Our algorithm | 33.1 | 0.090 |

### 3.4. Comparison of the computation time and memory usage

Finally, we compared the computation time and memory usage of *PoreSpy* and our algorithm (Table 3). The measurements were performed on a workstation equipped with a 56-core Intel(R) Xeon(R) Gold 6348 CPU. The time required for particle placement was defined as the "structure generation time," corresponding to the process up to Fig. 1(d) in our algorithm. In this case, *PoreSpy* required approximately 2,000 seconds, whereas our algorithm required only about 600 s. The time required for porosity adjustment was defined as the "effective porosity adjustment time," corresponding to the process from Fig. 1(d) to Fig. 1(h) in our algorithm, which took 869 s. Thus, our algorithm generated maps with a target porosity within 1% accuracy in a total of about 1500 s. In contrast, *PoreSpy* does not provide a porosity adjustment function, and therefore multiple map generations must be performed from the beginning. As a result, the effective porosity adjustment time in *PoreSpy* is approximately 2000 s multiplied by the number of trials. Regarding memory usage, *PoreSpy* consumed approximately three times more memory than our algorithm. *PoreSpy* allows particle overlaps, leading to an increase of the number of particles and memory consumption. In contrast, our algorithm imposes a condition on $\phi_{Particle}$ for particle placement (Fig. 1(c)). This reduces the number of particles and requires less memory compared to *PoreSpy*. These results demonstrate that our algorithm is more efficient in terms of both computation time and memory usage.

Table 3 Comparison of computational time and memory usage for map generation using *PoreSpy* and our algorithm.

|  | Structure generation time [sec] | Effective porosity adjustment time [sec] | Memory usage [GB] |
|---|---|---|---|
| *PoreSpy* | 2067 | – | 14 |
| Our algorithm | 595 | 869 | 5 |

### 4 Summary and Conclusions

In this study, we proposed a porous structure generation algorithm different from *PoreSpy*. Our algorithm enabled to generate porous structure with uniform density distribution and the porosity value that is very close to the input porosity value, resolving the issues in the *PoreSpy* program suite. In addition, it requires less computation time than *PoreSpy* due to parallelization, and it also achieves lower memory consumption. When generating *PoreSpy* and our algorithm with the parameters with which the maps obtained using the X-ray ptychography experiment, the error between the target porosity $\eta_{\text{Box}}$ and the effective porosity was approximately 4.3% for *PoreSpy*, whereas our algorithm achieved it within 1%. This error had a significant impact on the gas diffusion coefficient: the relative error of *PoreSpy* compared to experimental diffusion data exceeded 20%, while our method reproduced it within 5%. Therefore, our algorithm provides an efficient and high-precision alternative to *PoreSpy* for generating porous structures.

While our present algorithm is limited to the spherical particles, generation of porous structures with non-spherical or anisotropic particles (e.g., fibers or complex polymer morphologies) is also critical for understanding the gas diffusion layers in polymer electrolyte fuel cells and the filler dispersion structures in polymer composites. Extending our algorithm to handle such complex structures constitutes our primary focus for future work.

**Acknowledgment**

This study was financially supported by the JST (CREST Grant Number JPMJCR2233), Japan.

**Conflict of Interest**

There are no conflicts of interest to declare.

**Author Contributions**

S.A. conceived the idea, performed all computations and analyzed the results; All authors discussed the results, wrote the manuscript, and commented on the manuscript.

**Data Avairability**

Data will be made available on request.

**Code availability**

Our program, named "PorousGen," is available under the MIT License from https://github.com/YoshidomeGroup-Hydration/PorousGen. The use of the program is described on the webpage described above.